\documentclass[aps,prl,twocolumn,superscriptaddress,showpacs]{revtex4}
\usepackage{graphicx}

\begin{document}

\newcommand{\CP}{\mbox{\it CP}}
\newcommand{\ra}{\mbox{~$\rightarrow$}~}
\newcommand{\D}{\displaystyle}
\newcommand{\T}{\textstyle}
\newcommand{\GeVc}{\mbox{GeV/$c$}}
\newcommand{\GeVcc}{\mbox{GeV/$c^2$}}

\newcommand{\gL}{\mbox{$\Lambda$}}
\newcommand{\agL}{\mbox{$\overline{\Lambda}$}}
\newcommand{\Lz}{\mbox{$\Lambda^{\circ}$}}
\newcommand{\aLz}{\mbox{$\overline{\Lambda}$$^{\circ}$}}
\newcommand{\gXi}{\mbox{$\Xi$}}
\newcommand{\agXi}{\mbox{$\overline{\Xi}$}}
\newcommand{\Xipm}{\mbox{$\Xi^{\pm}$}}
\newcommand{\Xim}{\mbox{$\Xi^-$}}
\newcommand{\aXim}{\mbox{$\overline{\Xi}$$^+$}}
\newcommand{\gOm}{\mbox{$\Omega$}}
\newcommand{\agOm}{\mbox{$\overline{\Omega}$}}
\newcommand{\Omm}{\mbox{$\Omega^-$}}
\newcommand{\Ompm}{\mbox{$\Omega^{\pm}$}}
\newcommand{\aOmm}{\mbox{$\overline{\Omega}$$^+$}}

\newcommand{\gp}{\mbox{$p$}}
\newcommand{\agp}{\mbox{$\overline{p}$}}
\newcommand{\gppm}{\mbox{$p^{\pm}$}}
\newcommand{\gpmp}{\mbox{$p^{\mp}$}}

\newcommand{\piz}{\mbox{$\pi^{\circ}$}}
\newcommand{\gpi}{\mbox{$\pi$}}
\newcommand{\pip}{\mbox{$\pi^+$}}
\newcommand{\pim}{\mbox{$\pi^-$}}
\newcommand{\pipm}{\mbox{$\pi^{\pm}$}}
\newcommand{\pimp}{\mbox{$\pi^{\mp}$}}

\newcommand{\Kz}{\mbox{$K^0$}}
\newcommand{\Kl}{\mbox{$K_{L}$}}
\newcommand{\Klz}{\mbox{$K_{L}^0$}}
\newcommand{\Ksz}{\mbox{$K_{S}^0$}}
\newcommand{\Kpm}{\mbox{$K^{\pm}$}}
\newcommand{\Bd}{\mbox{$B_d$}}

\newcommand{\galpha}{\mbox{$\alpha$}}
\newcommand{\agalpha}{\mbox{$\overline{\alpha}$}}
\newcommand{\gbeta}{\mbox{$\beta$}}
\newcommand{\agbeta}{\mbox{$\overline{\beta}$}}
\newcommand{\ggamma}{\mbox{$\gamma$}}
\newcommand{\aggamma}{\mbox{$\overline{\gamma}$}}
\newcommand{\alp}{\makebox{$\alpha_p$}}
\newcommand{\bp}{\makebox{$\beta_p$}}
\newcommand{\gap}{\makebox{$\gamma_p$}}
\newcommand{\alal}{\makebox{$\alpha\alpha$}}
\newcommand{\alalbar}{\makebox{$\overline{\alpha}\overline{\alpha}$}}
\newcommand{\delalal}{\makebox{$\delta\alpha\alpha$}}
\newcommand{\delalalbar}{\makebox{$\delta\overline{\alpha}\overline{\alpha}$}}
\newcommand{\alXi}{\makebox{$\alpha_{\Xi}$}}
\newcommand{\aalXi}{\makebox{$\alpha_{\overline{\Xi}}$}}
\newcommand{\aalXibig}{\makebox{$\overline{\alpha}_{\Xi}$}}
\newcommand{\bXi}{\makebox{$\beta_{\Xi}$}}
\newcommand{\gaXi}{\makebox{$\gamma_{\Xi}$}}
\newcommand{\gamXi}{\makebox{$\gamma_{\Xi}$}}
\newcommand{\alL}{\makebox{$\alpha_{\Lambda}$}}
\newcommand{\aalL}{\makebox{$\alpha_{\overline{\Lambda}}$}}
\newcommand{\aalLbig}{\makebox{$\overline{\alpha}_{\Lambda}$}}
\newcommand{\alOm}{\makebox{$\alpha_{\Omega}$}}
\newcommand{\aalOm}{\makebox{$\alpha_{\overline{\Omega}}$}}
\newcommand{\aalOmbig}{\makebox{$\overline{\alpha}_{\Omega}$}}
\newcommand{\betaOm}{\makebox{$\beta_{\Omega}$}}
\newcommand{\delaL}{\makebox{$\Delta\alpha_{\Lambda}$}}
\newcommand{\delaXi}{\makebox{$\Delta\alpha_{\Xi}$}}

\newcommand{\AXi}{\makebox{$A_{\Xi}$}}
\newcommand{\aAXi}{\makebox{$A_{\overline{\Xi}}$}}
\newcommand{\Aratio}{\makebox{$\frac{\alpha + \overline{\alpha}}%
                                    {\alpha - \overline{\alpha}}$}}
\newcommand{\Aratiobig}{\makebox{$\D\frac{\alpha + \overline{\alpha}}%
                                         {\alpha - \overline{\alpha}}$}}
\newcommand{\AXiratio}{\makebox{$\frac{\alXi + \aalXi}{\alXi - \aalXi}$}}
\newcommand{\AXiratiobig}{\makebox{$\D\frac{\alXi + \aalXibig}{\alXi - \aalXibig
}$}}
\newcommand{\AXiratiotxt}{\makebox{$(\alXi + \aalXi)/(\alXi - \aalXi)$}}
\newcommand{\AXiratiobigtxt}{\makebox{$(\alXi+\aalXibig)/(\alXi-\aalXibig)$}}
\newcommand{\AL}{\makebox{$A_{\Lambda}$}}
\newcommand{\aAL}{\makebox{$A_{\overline{\Lambda}}$}}
\newcommand{\ALratio}{\makebox{$\frac{\alL + \aalL}{\alL - \aalL}$}}
\newcommand{\ALratiobig}{\makebox{$\D\frac{\alL + \aalLbig}{\alL - \aalLbig}$}}
\newcommand{\ALratiotxt}{\makebox{$(\alL + \aalL)/(\alL - \aalL)$}}
\newcommand{\ALratiobigtxt}{\makebox{$(\alL + \aalLbig)/(\alL - \aalLbig)$}}
\newcommand{\AXiL}{\makebox{${A}_{\Xi\Lambda}$}}
\newcommand{\delAXiL}{\makebox{$\Delta{A}_{\Xi\Lambda}$}}
\newcommand{\deldelta}{\makebox{$\Delta\delta$}}
\newcommand{\AXiLratio}{\makebox{$\frac{\alXi\alL - \aalXi\aalL}
                                       {\alXi\alL + \aalXi\aalL}$}}
\newcommand{\AXiLratiobig}{\makebox{$\D\frac{\alXi\alL - \aalXibig\aalLbig}
                                            {\alXi\alL + \aalXibig\aalLbig}$}}
\newcommand{\AXiLratiotxt}{\makebox{$(\alXi\alL-\aalXi\aalL)/
                                     (\alXi\alL+\aalXi\aalL)$}}
\newcommand{\AXiLratiotxtbig}{\makebox{$(\alXi\alL-\aalXibig\aalLbig)/
                                     (\alXi\alL+\aalXibig\aalLbig)$}}

\newcommand{\alphaLYt}{\makebox{$\alpha = {2\mbox{Re}(S^{\ast}P)}/(|S|^2 + |P|^2)$}}

\newcommand{\hypdkN}{\makebox{$\displaystyle
    \frac{dN}{d\Omega} =
    \frac{N_0}{4\pi}(1 + \alpha\vec{P}_p{\cdot}\hat{p}_d)$}}
\newcommand{\hypdkXiLN}{\makebox{$\displaystyle
    \frac{dN}{d\cos\theta} =
    \frac{N_0}{2}(1 + \alpha_{\Xi}\alpha_{\Lambda}\cos\theta)$}}

%\preprint{}

\title{Search for \textbf{\textit{CP}} Violation in 
       Charged-{\boldmath$\Xi$} and {\boldmath$\Lambda$} Hyperon Decays}

\affiliation{Institute of Physics, Academia Sinica, Taipei 11529, 
   Taiwan, Republic of China}
\affiliation{University of California, Berkeley, California 94720, USA}
\affiliation{Fermi National Accelerator Laboratory, Batavia, Illinois 
             60510, USA}
\affiliation{Universidad de Guanajuato, 37000 Le\'{o}n, Mexico}
\affiliation{Illinois Institute of Technology, Chicago, Illinois 60616, USA}
\affiliation{Universit\'{e} de Lausanne, CH-1015 Lausanne, Switzerland}
\affiliation{Lawrence Berkeley National Laboratory, Berkeley, California 
             94720, USA}
\affiliation{University of Michigan, Ann Arbor, Michigan 48109, USA}
\affiliation{University of South Alabama, Mobile, Alabama 36688, USA}
\affiliation{University of Virginia, Charlottesville, Virginia 22904, USA}

\author{T. Holmstrom}
\affiliation{University of Virginia, Charlottesville, Virginia 22904, USA}
\author{N. Leros}
\affiliation{Universit\'{e} de Lausanne, CH-1015 Lausanne, Switzerland}
\author{R.A. Burnstein}
\affiliation{Illinois Institute of Technology, Chicago, Illinois 60616, USA}
\author{A. Chakravorty}
\affiliation{Illinois Institute of Technology, Chicago, Illinois 60616, USA}
\author{A. Chan}
\affiliation{Institute of Physics, Academia Sinica, Taipei 11529, 
   Taiwan, Republic of China}
\author{Y.C. Chen}
\affiliation{Institute of Physics, Academia Sinica, Taipei 11529, 
   Taiwan, Republic of China}
\author{W.S. Choong}
\affiliation{University of California, Berkeley, California 94720, USA}
\affiliation{Lawrence Berkeley National Laboratory, Berkeley, California 
             94720, USA}
\author{K. Clark}
\affiliation{University of South Alabama, Mobile, Alabama 36688, USA}
\author{E.C. Dukes}
  \email[To whom correspondence should be addressed.  Electronic address: ]{craigdukes@virginia.edu.}
\affiliation{University of Virginia, Charlottesville, Virginia 22904, USA}
\author{C. Durandet}
\affiliation{University of Virginia, Charlottesville, Virginia 22904, USA}
\author{J. Felix}
\affiliation{Universidad de Guanajuato, 37000 Le\'{o}n, Mexico}
\author{Y. Fu}
\affiliation{Lawrence Berkeley National Laboratory, Berkeley, California 
             94720, USA}
\author{G. Gidal}
\affiliation{Lawrence Berkeley National Laboratory, Berkeley, California 
             94720, USA}
\author{P. Gu}
\affiliation{Lawrence Berkeley National Laboratory, Berkeley, California 
             94720, USA}
\author{H.R. Gustafson}
\affiliation{University of Michigan, Ann Arbor, Michigan 48109, USA}
\author{C. Ho}
\affiliation{Institute of Physics, Academia Sinica, Taipei 11529, 
   Taiwan, Republic of China}
\author{M. Huang}
\affiliation{University of Virginia, Charlottesville, Virginia 22904, USA}
\author{C. James}
\affiliation{Fermi National Accelerator Laboratory, Batavia, Illinois 
             60510, USA}
\author{C.M. Jenkins}
\affiliation{University of South Alabama, Mobile, Alabama 36688, USA}
\author{T. Jones}
\affiliation{Lawrence Berkeley National Laboratory, Berkeley, California 
             94720, USA}
\author{D.M. Kaplan}
\affiliation{Illinois Institute of Technology, Chicago, Illinois 60616, USA}
\author{L.M. Lederman}
\affiliation{Illinois Institute of Technology, Chicago, Illinois 60616, USA}
\author{M.J. Longo}
\affiliation{University of Michigan, Ann Arbor, Michigan 48109, USA}
\author{F. Lopez}
\affiliation{University of Michigan, Ann Arbor, Michigan 48109, USA}
\author{L.C. Lu}
\affiliation{University of Virginia, Charlottesville, Virginia 22904, USA}
\author{W. Luebke}
\affiliation{Illinois Institute of Technology, Chicago, Illinois 60616, USA}
\author{K.B. Luk}
\affiliation{University of California, Berkeley, California 94720, USA}
\affiliation{Lawrence Berkeley National Laboratory, Berkeley, California 
             94720, USA}
\author{K.S. Nelson}
\affiliation{University of Virginia, Charlottesville, Virginia 22904, USA}
\author{H.K. Park}
\affiliation{University of Michigan, Ann Arbor, Michigan 48109, USA}
\author{J.-P. Perroud}
\affiliation{Universit\'{e} de Lausanne, CH-1015 Lausanne, Switzerland}
\author{D. Rajaram}
\affiliation{Illinois Institute of Technology, Chicago, Illinois 60616, USA}
\author{H.A. Rubin}
\affiliation{Illinois Institute of Technology, Chicago, Illinois 60616, USA}
\author{P.K. Teng}
\affiliation{Institute of Physics, Academia Sinica, Taipei 11529, 
   Taiwan, Republic of China}
\author{J. Volk}
\affiliation{Fermi National Accelerator Laboratory, Batavia, Illinois
             60510, USA}
\author{C.G. White}
\affiliation{Illinois Institute of Technology, Chicago, Illinois 60616, USA}
\author{S.L. White}
\affiliation{Illinois Institute of Technology, Chicago, Illinois 60616, USA}
\author{P. Zyla}
\affiliation{Lawrence Berkeley National Laboratory, Berkeley, California
             94720, USA}

\collaboration{HyperCP Collaboration}
\noaffiliation

\date{December 13, 2004}

\begin{abstract}
We have compared the \gp\ and \agp\ angular distributions in 
117 million \Xim\ra\gL\pim\ra\gp\pim\pim\ and  41 million 
\aXim\ra\agL\pip\ra\agp\pip\pip\ decays using a subset of the data 
from the HyperCP experiment (E871) at Fermilab.
We find no evidence of \CP\ violation, with the direct-\CP-violating
parameter $\AXiL\equiv\AXiLratiotxtbig\ =
[0.0{\pm}5.1({\rm stat}){\pm}4.4({\rm syst})]{\times}10^{-4}$.

\end{abstract}

%\keywords{}

\maketitle

In the standard model, \CP\ asymmetries are expected to be ubiquitous 
in weak interaction processes, albeit often vanishingly small. 
To date, \CP\ asymmetries have been seen only in the decays of 
\Klz\ \cite{KDCP} and \Bd\ mesons \cite{BCP}.
Although the asymmetries observed in these decays
are consistent with standard-model predictions,
exotic sources of \CP\ violation have not been ruled out.
Hence it is vital to search for novel sources of \CP\ violation.
Hyperon decays offer promising possibilities for such searches
as they are sensitive to sources of \CP\ violation that, for example,
neutral kaon decays are not \cite{Deshpande,Tandean}.
The most experimentally accessible \CP-violating signature
in spin-$\frac{1}{2}$ hyperon decays is the difference between
hyperon and anti-hyperon decay distributions in
their parity-violating two-body weak decays.
In such decays the angular distribution of the daughter baryon is
$dN/d\Omega = N_0(1 + \alpha\vec{P}_p{\cdot}\hat{p}_d)/4\pi$,
where $\vec{P}_p$ is the parent polarization,
$\hat{p}_d$ is the daughter baryon direction,
and \alphaLYt, with $S$ and $P$ the $l = 0$ (parity-odd)
and $l = 1$ (parity-even) final-state amplitudes.
\CP\ invariance requires that $\alpha = -\overline{\alpha}$ \cite{Pais}.

In HyperCP, the \Xim\ and \aXim's were produced at an average angle 
of $0^{\circ}$ so that their polarization was zero.
The angular distribution of \gp's from unpolarized \Xim's in
\Xim\ra\gL\pim\ra\gp\pim\pim\ decays is given by
\begin{equation}
   \hypdkXiLN,
\label{eq:2}
\end{equation}
since the daughter \gL\ is produced in a helicity state with
polarization \alXi\ \cite{LeeYang}.
The polar angle $\theta$ is measured in that \gL\ rest frame, 
called the lambda helicity frame, in which the direction
of the \gL\ in the \Xim\ rest frame defines the polar axis.
The angular distribution of the \agp\ from the corresponding
decay sequence, \aXim \ra \agL\pip \ra \agp\pip\pip,
should be identical {\em if \CP\ is not violated},
as both \alXi\ and \alL\ reverse sign.
Any difference in the angular distributions is evidence of \CP\
violation in either \gXi\ or \gL\ decays, or perhaps both.
The measured \CP-violating observable is
\begin{equation}
  A_{\Xi\Lambda}\equiv\AXiLratiobig\approx\AXi + \AL,
\label{eq:3}
\end{equation}
where $\AXi\equiv\AXiratiobigtxt$ and $\AL\equiv\ALratiobigtxt$.

The most recent standard-model calculation for the combined asymmetry 
is $-0.5{\times}10^{-4}{\leq}\AXiL{\leq}0.5{\times}10^{-4}$ \cite{SM_calc}.
Note that this prediction uses a theoretical calculation of the 
$S$- and $P$-wave \gL\gpi\ final-state scattering phase-shift differences
rather than more recent measurements \cite{XiLps}.
Non-standard-model calculations, such as left-right symmetric
models \cite{LR_calc} and supersymmetric models \cite{SUSY_He,SUSY_Chen}, 
allow for much larger asymmetries.
The supersymmetric calculation of He {\em et al.}\ \cite{SUSY_He}
generates values of \AL\ as large as $19{\times}10^{-4}$.
Bounds from $\epsilon$ and $\epsilon^{\prime}/\epsilon$ in \Kz\ decays
limit \AXiL\ to be less than $97{\times}10^{-4}$ \cite{Tandean}.
Experiments have yet to probe hyperon \CP\ asymmetries beyond the
$O(10^{-2})$ level, with the best limit being $\AXiL = +0.012{\pm}0.014$
\cite{E756}.  In this Letter we present an experimental search with
significantly improved sensitivity.

\begin{figure}[htb]
\includegraphics[width=3.4in]{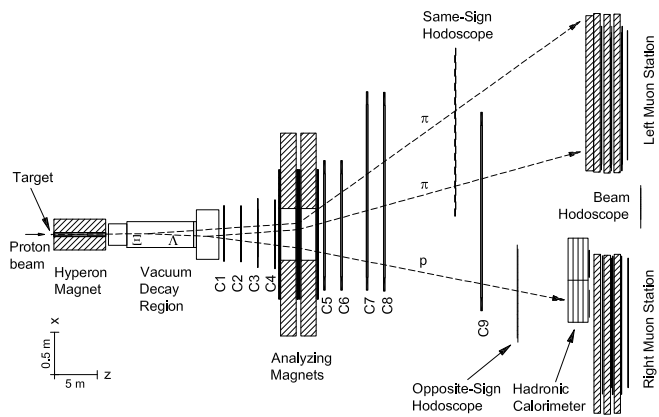}
\caption{Plan view of the HyperCP spectrometer.
   \label{fig:spect_plan}}
\end{figure}

Data were taken at Fermilab using a high-rate spectrometer 
(Fig.~\ref{fig:spect_plan}) \cite{NIM_spect}. 
The hyperons were produced by an 800~GeV/$c$ proton beam incident
at $0^{\circ}$ on a $2{\times}2$\,mm$^2$ Cu target.
Immediately after the target was a 6.096\,m long curved collimator 
embedded in a dipole magnet (``hyperon magnet'').  Charged particles 
following the central orbit of the collimator exited
upward at 19.51~mrad to the incident 
proton beam direction with a momentum of 157~GeV/$c$.
Following the collimator was a 13\,m long evacuated pipe 
(``vacuum decay region'').
The momenta of charged particles were measured
using nine multiwire proportional chambers (MWPCs), 
four in front and five behind two dipole magnets (``analyzing magnets'').
At the rear of the spectrometer were two scintillator hodoscopes
used in the trigger:  one, the same-sign (SS) hodoscope,
situated to the beam-left of the charged secondary beam, 
the other, the opposite-sign (OS) hodoscope, situated to beam-right.
A hadronic calorimeter was used to trigger on the energy of the \gp\ or \agp.

The \Xim\ (negative) and \aXim\ (positive) data were taken alternately,
typically with three positive runs followed by one negative run in a 
sequence that usually took about 12 hours.
To switch from one running mode to the other, the polarities of 
the hyperon and analyzing magnets were reversed and the targets were
interchanged; differing target lengths were used to keep the 
secondary-beam rates approximately equal.
At a nominal primary proton beam rate of $7.5{\times}10^{9}$\,s$^{-1}$
the secondary-beam rate was $13{\times}10^6$\,s$^{-1}$, with
the average difference between the positive and negative rates
less than 5\%.
A simple trigger --- the ``cascade'' (CAS) trigger --- with large acceptance 
and single-bucket (18.9~ns) time resolution was used to select 
events with the \gL \ra \gp\pim\ topology.
It required the coincidence of at least one hit in
each of the SS and OS hodoscopes --- the ``left-right'' (LR) subtrigger ---
along with at least ${\approx}$\,40\,GeV energy deposited 
in the hadronic calorimeter, 
an amount well below that of the lowest energy \gp\ or \agp.

A total of 90 billion CAS triggers were recorded
in the 1999--2000 running period.
This analysis used data taken from the end of the run:
a 21-day period in December, 1999 and a 12-day period in January, 2000.
(The intervening period was devoted to special polarized \gXi\ runs.)
The dataset included 19\% of all the good \aXim\ events (41.4 million)
and 14\% of all the good \Xim\ events (117.3 million) taken in the 
1999--2000 running period.
The data were divided into 18 analysis sets
of roughly equal size, each containing at least three
positive and one negative run taken closely spaced in time.
\begin{figure}[b]
\includegraphics[width=2.2in]{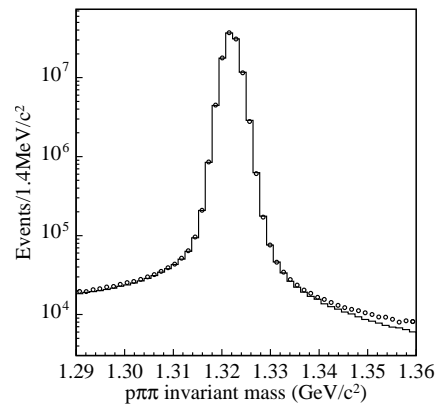}
\caption{The unweighted $\gp\pim\pim$ (histogram) and $\agp\pip\pip$ (circles) 
         invariant masses.
   \label{fig:xi_masses}}
\end{figure}

The data were analyzed by a computer program that reconstructed tracks
and determined particle momenta, invariant masses, and decay vertices, 
assuming the \gXi\ra\gL\gpi\ and \gL\ra\gp\gpi\ hypotheses.
Efficiencies of each MWPC wire and hodoscope counter
were measured on a run-by-run basis using tracks from reconstructed 
\Xipm, \Kpm, and \Ompm\ events.  These efficiencies were typically
${\approx}\,99$\% and ${>}\,99$\% for the MWPCs and hodoscopes, respectively.
The calorimeter trigger efficiency, as determined on a run-by-run basis
using good \Xipm\ events from the LR trigger, was ${>}\,99$\%.
Runs with anomalously low (${\lesssim}\,95$\%) hodoscope, wire chamber, 
or calorimeter efficiencies were not used; these were less than 5\% 
of the total.  The criteria used to select the final event samples were:
(1) that the \gp\gpi\ and \gp\gpi\gpi\ 
   invariant masses be, respectively, within ${\pm}5.6$~MeV/$c^2$ 
   ($3.5\,\sigma$) and ${\pm}3.5$~MeV/$c^2$ ($3.5\,\sigma$) of the 
   mean values of the \gXi\ and \gL\ masses (1.3220 and 1.1158\,GeV/$c^2$);
(2) that the $z$ coordinate of the \gXi\ and \gL\ decay vertices lie within
   the vacuum decay region 
   and that the \gL\ decay vertex precede the \gXi\ decay vertex by 
   no more than 0.50~m;
(3) that the reconstructed \gXi\ trajectory trace back to 
   within ${\pm}2.45$~mm ($3.3\,\sigma$) and ${\pm}3.26$\,mm ($3.4\,\sigma$), 
   respectively, in $x$ and $y$, of the center of the target;
(4) that the \gXi\ trajectory trace back to within
   $+8.2/-8.4$\,mm and ${\pm}6.5$\,mm, respectively in $x$ and $y$,
   from the center of the exit of the collimator; and
(5) that the \pipm\pipm\pimp\ invariant mass be greater than
   0.5\,GeV/$c^2$ (to remove \Kpm\ra\pipm\pipm\pimp\ decays).
Cuts on the particle momenta and
the numbers of SS and OS hodoscope hits were also made.
Events satisfying these criteria had a background to signal ratio
of $(0.43{\pm}0.03)$\% for the \Xim\ data and $(0.41{\pm}0.03)$\% for the 
\aXim\ data (Fig.~\ref{fig:xi_masses}).

The \CP\ asymmetry \AXiL\ was extracted by comparing the \gp\ and 
\agp\ $\cos\theta$ distributions in the lambda helicity frame.
As care was taken to exactly reverse the hyperon
and analyzing magnetic fields --- the fractional difference between
the magnitudes of positive and negative analyzing magnet fields was
${\approx}\,3{\times}10^{-4}$ --- 
biases due to spatial acceptance differences were minimal.
The magnetic field magnitudes were updated on a spill-by-spill basis
using values recorded by Hall probes placed in each magnet.
Differences in the MWPC wire efficiencies were typically
on the order of $1{\times}10^{-3}$ in the secondary beam region,
and much less outside.
Hodoscope counter efficiency differences were typically much
less than $1{\times}10^{-3}$.
These efficiency differences had negligible effects on \AXiL.
The calorimeter efficiency difference was ${\approx}\,1{\times}10^{-3}$,
and, within errors, uniform over the calorimeter face.

To eliminate differences in the \Xim\ and \aXim\ momentum and position 
distributions, the \Xim\ and \aXim\
events were weighted in the three momentum-dependent parameters
of the \gXi's at the collimator exit (their effective production point):  
the momentum ($p_{\Xi}$), the $y$ coordinate ($y_{\Xi}$), and
the $y$ slope ($s_{{\Xi}y}$).
Each parameter was binned in 100 bins for a total of $10^6$ bins.
The $p_{\Xi}$, $y_{\Xi}$, and $s_{{\Xi}y}$ bin widths were, respectively,
2.25~GeV/$c$, 0.13~mm, and $0.08{\times}10^{-3}$.
Bins with fewer than four events of either polarity had their
weights set to zero.
After the weights were computed the \gp\ (or \agp) $\cos\theta$
of each event was weighted appropriately and the ratio of the weighted 
\gp\ and \agp\ $\cos\theta$ distributions was then formed.  
The expected ratio,
\begin{equation}
   R = C\frac{1 + \alXi\alL\cos\theta}
             {1 + (\alXi\alL - \delta)\cos\theta},
\label{eq:4}
\end{equation}
determined using Eq.~(\ref{eq:2}),
was fit to the data to extract the asymmetry
$\delta \equiv \alXi\alL - \aalXibig\aalLbig \cong 2\alXi\alL{\cdot}\AXiL$ 
and the scale factor $C$,
where $\alXi\alL = -0.294$ \cite{PDG} was used.
No acceptance or efficiency corrections were made.
\begin{figure}[b]
\includegraphics[width=3.3in]{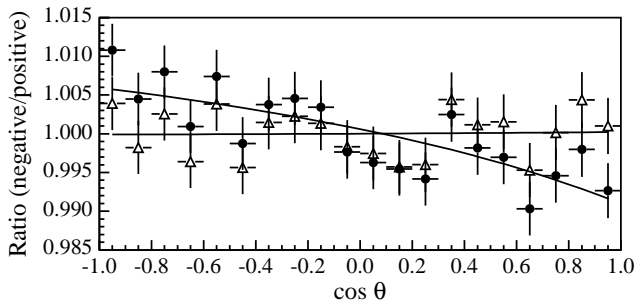}
\caption{Ratios of \gp\ to \agp\ $\cos\theta$ distributions
         from analysis set 1, both unweighted (filled circles) 
         and  weighted (open triangles), with fits to the form
         given in Eq.~(\ref{eq:4}).
   \label{fig:cos_theta}}
\end{figure}

\begin{figure}[b]
\includegraphics[width=3.3in]{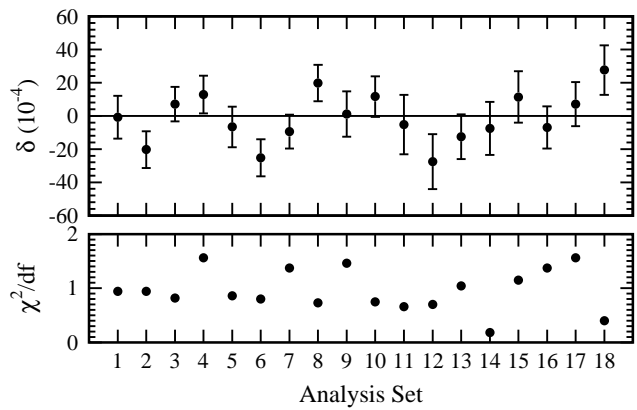}
\caption{Asymmetry $\delta$ (top) and $\chi^2$/df (bottom) versus Analysis Set.
   \label{fig:results}}
\end{figure}

Figure~\ref{fig:cos_theta} shows typical ratios of weighted and unweighted
\gp\ to \agp\ $\cos\theta$ distributions from analysis set 1.
Figure~\ref{fig:results} shows $\delta$ for all 18 analysis sets.
Fits to $R$ were good; the average $\chi^2/{\rm df}$ was 0.96.
The weighted average of $\delta$ for all 18 analysis sets is
$\delta = (-1.3{\pm}3.0){\times}10^{-4}$, where the
error is statistical, with $\chi^2 = 24$.  
The corresponding raw asymmetry is 
$\AXiL({\rm raw}) = (2.2{\pm}5.1){\times}10^{-4}$.

The background-corrected asymmetry was determined as follows.
The asymmetries in the mass sidebands 1.290 -- 1.310~\GeVcc\ and
1.334 -- 1.354~\GeVcc\ were found, using weights from the central region.
The weighted average of the two sideband asymmetries, 
scaled by the average background fraction of 0.42\%,
was subtracted from the raw asymmetry to give 
$\AXiL = (0.0{\pm}5.1){\times}10^{-4}$.

The analysis algorithm and its implementation were verified by a
simulation, called the collimator hybrid Monte Carlo (CHMC),
that used momenta and positions at the collimator exit from real
\Xim\ and \aXim\ events as input to computer-generated \gXi\ decays.
Using zero and near-zero input asymmetries, the extracted values
of \AXiL\ differed from the input values by $(-1.9{\pm}1.6){\times}10^{-4}$.
Note that the measurement of \AXiL\ has no Monte Carlo dependence.

Systematic errors were small for several reasons.
First, common biases were suppressed by
taking the ratio of the \gp\ and \agp\ $\cos\theta$ distributions.
Second, overall efficiency differences do not cause a bias,
only spatially dependent differences.
Finally, since the polar  axis changes from event to event
in the lambda helicity frame, there is only a weak correlation 
between any particular region of the apparatus and 
$\theta$, minimizing biases due
to localized differences in detector efficiencies.
Table~\ref{tab:systematics} lists the systematic errors;
added in quadrature they give $4.4{\times}10^{-4}$.
\begin{table}[htb]
\renewcommand{\arraystretch}{0.25}
\caption{Systematic errors.\label{tab:systematics}}
\begin{tabular}{lr@{.}l}
  \hline\hline
  \multicolumn{3}{l}{\hspace*{1.1in}Source\hspace*{1.0in}Error ($10^{-4}$)} \\ 
  \hline
  Analyzing magnets field uncertainties \hspace*{0.6in} &  2 & 4  \\
  Calorimeter inefficiency uncertainty          &  2 & 1  \\
  Validation of analysis code                   &  1 & 9  \\
  Collimator exit $x$ slope cut                 &  1 & 4  \\
  Collimator exit $x$ position cut              &  1 & 2  \\
  MWPC inefficiency uncertainty                 &  1 & 0  \\
  Hodoscope inefficiency uncertainty            &  0 & 3  \\
  Particle/antiparticle interaction differences &  0 & 9  \\
  Momentum bin size                             &  0 & 4  \\
  Background subtraction uncertainty            &  0 & 3  \\
  Error on \alXi\alL\                           &  0 & 03 \\[0.05in]
\hline\hline
\end{tabular}
\end{table}

The largest systematic error was due to the uncertainty in
the calibrations of the Hall probes in the analyzing magnets.  
The magnetic fields were quite stable:
variations in the Hall probe readings of the sum
of the two fields were 6.3\,G and 5.7\,G (rms), respectively, for the 
9024 positive and 2396 negative-polarity spills used in this analysis.
From calibrations before and after
the running periods using more precise NMR probes, 
the uncertainty was estimated to be 5.5\,G for the sum of the fields,
corresponding to an uncertainty of $2.4{\times}10^{-4}$ in \AXiL.

The differences in efficiencies of the calorimeter, 
hodoscopes, and MWPCs between positive and negative running,
were not corrected for, as they were negligibly small.
The effect of calorimeter inefficiency differences was
determined using a data sample taken with the LR trigger.
The difference in \AXiL, with and without the calorimeter
trigger requirement, was found to be consistent with zero, 
with a statistical error of $2.1{\times}10^{-4}$.
Weighting events to correct for the hodoscope inefficiencies
changed \AXiL\ by only $0.3{\times}10^{-4}$.
The effect of MWPC inefficiency differences 
($1.0{\times}10^{-4}$) was estimated using CHMC data by determining 
the difference in \AXiL\ using real and 100\% efficiencies.

The effect on \AXiL\ of tighter cuts on the (unweighted) \gXi\ $x$ 
slope and position at the collimator exit was studied 
and resulted in respective uncertainties of 
$1.4{\times}10^{-4}$ and $1.2{\times}10^{-4}$.
The effect of the bin sizes used in extracting the event weights
was investigated by increasing and decreasing the \gXi\ momentum
bin sizes by 25\%,
\AXiL\ being most sensitive to momentum.
Another possible source of bias was a momentum-dependent differential loss 
of events due to interactions of the \Xim\ and \aXim\ decay products
with material in the spectrometer.
Monte Carlo studies, using the interaction cross sections given in 
Ref.~\cite{PDG}, showed this bias was negligible.

The result was stable with respect to time,
\gXi\ momentum, and secondary-beam intensity.
Differences between the \Xim\ and \aXim\ production angles 
could in principle cause a bias due to production polarization
differences.  Average production angle
differences were only ${\approx}\,{0.02}$~mrad.  
Assuming a linear dependence 
of the polarization on transverse momentum \cite{PHo}, 
Monte Carlo studies indicated a 
negligible effect on the \gp\ and \agp\ $\cos\theta$ slopes.
No dependence of \AXiL\ on production angle or incident proton beam
position was evident.

To conclude, we have measured \AXiL\ to be
$[0.0{\pm}5.1({\rm stat}){\pm}4.4({\rm syst})]{\times}10^{-4}$.
This result is consistent with standard-model predictions
and is a factor of 20 improvement over the best previous result
\cite{E756}.

The authors are indebted to the staffs of Fermilab and the
participating institutions for their vital contributions.
This work was supported by the U.S. Department of Energy
and the National Science Council of Taiwan, R.O.C\@.
E.C.D. and K.S.N. were partially supported by the 
Institute for Nuclear and Particle Physics.
K.B.L. was partially supported by the Miller Institute.

\end{document}